\documentclass[11pt]{article}
\usepackage{graphicx}

\begin{document}
\title{Entanglement and Adiabatic Quantum Computation}
\author{Daria Ahrensmeier\footnote{based on a talk by DA given at Theory Canada I, UBC, Vancouver, June 2005}\\[0.2cm]
 Department of Physics and Winnipeg Institute for Theoretical Physics,\\
 University of Winnipeg,\\
  Winnipeg, MB R3B 2E9, Canada\\[0.1cm]
  d.ahrensmeier@uwinnipeg.ca}
\maketitle
\begin{abstract}
Adiabatic quantum computation provides an alternative approach to quantum computation using a time-dependent Hamiltonian. The time evolution of entanglement during the adiabatic quantum search algorithm is studied, and its relevance as a resource is discussed.
\end{abstract}
%

\def\tablefootnote#1{%
\hbox to \textwidth{\hss\vbox{\hsize\captionwidth\footnotesize#1}\hss}} 

\section{Introduction}
How does quantum computation work? What are the necessary physical resources? And what can we learn from it about fundamental physics? Adiabatic quantum computation may provide some contributions to these questions.

Quantum information science has gone through an amazing development in the past decade, and fascinating applications have been developed, including quantum teleportation or quantum cryptography. It is hoped  that quantum computing will provide improved performance as compared to a classical computer, that it will provide an alternative route where miniaturization reaches a limit, and that it might even allow to solve problems that a classical computer cannot solve.  But despite the amazing success of quantum information science with regard to applications, there is still no general agreement about the fundamental physics principles or physical resources that make quantum computing (at least theoretically) work and that provide a speed-up over classical computation. 
Candidates are entanglement or nonlocality, or the Hilbert space structure including the superposition principle, which suggests that some mechanism of parallel computing might be at work.

In this paper, I study the question of resources within an alternative model of quantum computation, so called adiabatic quantum computation (see, e.g., \cite{AQC}). The equivalence of adiabatic and standard model quantum computation has recently been shown in \cite{Aharonov}. 
The advantage for the investigation of entanglement is that in the adiabatic approach, the computation is done through  a time-dependent Hamiltonian, which allows to study the evolution of physical quantities like the entropy of entanglement as a function of time. 
Below, I will provide a brief introduction to entanglement and to adiabatic quantum computation and its realization of the search algorithm. I will describe a method for calculating entanglement during the running time of the algorithm and present results for the original algorithm as well as modifications. Note that the presentation is aimed at a physics audience and does not require previous knowledge of quantum computation.

\section{Physical Resources in quantum information science}

The subfield of quantum information science that has evolved most rapidly and already seems to be moving from purely scientific research to industrial applications (e.g.,\cite{poppe0404}) is quantum cryptography. In this case, as in the related quantum teleportation (see,e.g.,\cite{Nielsen} for an illuminating discussion), entanglement is used as a resource, and the measurement process is part of the protocol. 

In comparison to these areas, the progress in quantum computation has been more modest. Algorithms have been found for the testing of functions (Deutsch-Josza's algorithm \cite{Deutsch}), for factorizing large integers in polynomial time (Shor's algorithm \cite{Shor}), and for searching (Grover's algorithm \cite{Grover}). Although the efficiency of these algorithms is well established, the question of which physical resources allow them to work is still a controversial issue (see e.g. \cite{Braunstein} for a discussion about possible implementations of these algorithms using Nuclear Magnetic Resonance and the question whether entanglement has been involved). It is generally believed that entanglement is necessary for quantum computing as a nonclassical feature providing the computational speedup \cite{Ekert}. It has also been suggested to consider this resource not as an intrinsic property of quantum physics itself, but something that is tied to a particular choice of mathematical formalism \cite{Josza}. On the other hand, examples have been presented where quantum computing without entanglement is better than the classical version \cite{Biham}. The search algorithm has even been implemented using the wave nature of classical Fourier optics \cite{Fourier}. Summarizing, it seems that entanglement somehow plays a role in quantum algorithms. But it is not understood yet whether the entanglement is necessary for the speed-up, or whether it is created by the algorithm as a by-product, if this depends on the algorithm considered or if it is a general property of quantum computing, and how this is related to the Hilbert space structure or the formalism in general and other physical resources involved, like energy.

\subsection{Entanglement}

Entanglement as a non-classical physical phenomenon (non-classical correlations) has been known since the early days of quantum mechanics. It plays the leading role in the  EPR paradox, although in that context, it is better known as non-locality -- and was nick-named ''spooky actions at a distance''.  When Bell established his famous inequalities, the discussion was moved from more philosophical considerations to experimentally testable physics. And yet, after all this time, a precise definition and especially a general measure for entanglement is still missing. 
Consider, for example, one of the Bell states, $1/\sqrt{2}(|00\rangle+|11\rangle)$. It cannot be written as a product of single qubit states, in other words, it is not separable. Unfortunately, this version of stating entanglement is not very practical in applications. It is desirable to have a measure of entanglement that is easy to calculate, and that allows to compare different states in their degree of entanglement. For bipartite systems (i.e. systems consisting of two subsystems, A and B)  in pure states, the von Neumann entropy of either of the subsystems is generally accepted as the measure of entanglement. It gives the amount of information of one qubit that can be obtained by making a measurement on the other qubit of a pair, or, in other words, it measures how complete the information about the subsystem is, using the density operators of the subsystems:
\begin{equation}
E(\rho_{AB}) =  S(\rho_A)=-Tr(\rho_A \log\rho_A) =  S(\rho_B)=-Tr(\rho_B \log\rho_B)
\end{equation}
with the traces over either subsystem
$\rho_A  =  Tr_B(\rho_{AB})$ and $\rho_B  =  Tr_A(\rho_{AB})$.
 For a system in an entangled state, the state of the whole system is completely known, but the state of each subsystem is not, i.e. it is in a mixed state. For a maximally entangled state of two qubits, e.g. a Bell state, $\rho_A=1/2 I$, and $E(\rho_{AB})=1$. If the subsystem is in a pure state, the entropy of entanglement for the composite system vanishes. 
The von Neumann entropy can be calculated from the (non-zero) eigenvalues $\lambda_n$ of the reduced density matrix as
\begin{equation}
 S  =  -Tr(\rho\log\rho) = - \sum_n \lambda_n \log \lambda_n.
\end{equation}
For more general systems, for example for mixed states, several other, inequivalent  measures of entanglement have been suggested, but no generally accepted one has been found yet (see, e.g., \cite{Vedral} for a discussion).

\section{Adiabatic Quantum Computation}

The building blocks of a quantum computer are qubits, quantum mechanical systems with a two-dimensional Hilbert space, which can be realized as, e.g., spin 1/2 particles. (Note that the general state of a qubit is a superposition of the two basis states, $|\Phi\rangle =\alpha |0\rangle + \beta |1\rangle$, whereas a classical bit is either in state 0 or state 1.)  In the traditional model of quantum computation, quantum circuits are considered which consist of a series of unitary operators (gates) acting on single or multiple qubits. At the end of the computation, a measurement provides the result.

The idea behind adiabatic quantum computation is to use a more ''natural'' description for the time evolution of the system, a Schr\"{o}dinger equation. The computation is understood as a continuous time evolution from an initial to a desired final state:
The initial state $|\Psi_0\rangle$ is the ground state of a Hamiltonian $H_0$ and (assumed to be) easy to build, whereas the final state $|\Psi_1\rangle$ is the ground state of a Hamiltonian $H_1$ and encodes the solution of the computational problem. The time dependent Hamiltonian drives  the system from the initial to the final instantaneous ground state. It is usually constructed as a linear combination
\begin{equation} 
H(t)=f(t)H_0 +g(t)H_1
\end{equation}
with the monotonic functions $f,g$ fulfilling
$f(0)=1, f(T)=0$ and $g(0)=0, g(T)=1$.
The system is supposed to stay in  its instantaneous ground state during this time evolution. The condition for this behaviour is provided by the adiabatic theorem:  Consider a system evolving according to a Schr\"{o}dinger equation with a time dependent Hamiltonian. Its instantaneous energy eigenstates obey
\begin{equation}H(t)|E_k;t\rangle=E_k(t)|E_k;t\rangle .
\end{equation}
If the system is prepared in its ground state $|E_0;0\rangle$ at $t=0$, after time $T$ the probability for staying in the instantaneous ground state (up to a phase) is
$|\langle E_0;T|\Psi(T)\rangle|^2 \geq 1-\epsilon^2$,
provided that the Hamiltonian varies slowly enough,
\begin{equation}\label{condition}
\frac{\max_{0\leq t\leq T}\left |\left \langle E_1;t\left |\frac{dH}
{dt}\right |E_0;t\right \rangle \right |}
{g_{min}^2}\leq \epsilon,
\end{equation}
with the minimum gap between the two lowest eigenvalues
\begin{equation}
g_{min}=\min_{0\leq t\leq T}[E_1(t)-E_0(t)].
\end{equation}
Accordingly, the running time of the algorithm is given by the minimum time $T_{min}$ for the evolution of the system to be adiabatic, eq. (\ref{condition}). After this time, a measurement will give the desired solution with probability $P \approx 1$, i.e. with accuracy $\epsilon$ given by eq.(\ref{condition}).

\section{Adiabatic Quantum Search}

The search problem consists  of finding a marked object (the ''needle'') in an unsorted database (the ''haystack'') of $N$ elements with the minimum amount of computational work. A classical search would simply examine the items one by one and find the marked one after $O(N)$ steps.
Grover's original algorithm \cite{Grover} allows to search an unstructured database quadratically faster than any classical algorithm. 
In the quantum version, the items of the database are represented by the computational basis states, and the haystack is given by an equally weighted superposition of these states,
\begin{equation}
 |\Psi_0\rangle = \frac{1}{\sqrt{N}}\sum_{i=0}^{N-1}|i\rangle ,
\end{equation}
i.e. each one has the same probability.
Grover's algorithm works by increasing the amplitude of the marked element while decreasing all other amplitudes, i.e. by rotating the state of the system towards the marked state.  Its action on a state $|\Psi\rangle$ can be written as
\begin{equation}
G|\Psi\rangle=(2|\Psi\rangle\langle\Psi|-I)(I-2|m\rangle\langle m|)|\Psi\rangle
    =(1-\frac{1}{N/4})|\Psi\rangle+\frac{2}{\sqrt{N}}|m\rangle .
\end{equation}
Note that the factor $1/\sqrt{N}$ is the overlap of the state $|\Psi\rangle$ with the marked state $|m\rangle$. After $k_0\approx \frac{\pi}{4}\sqrt{N}$ iterations, the probability for obtaining the marked state after a measurement is $P\geq 1-1/N$, i.e. the complexity of Grover's algorithm is $O(\sqrt{N})$, a quadratic improvement over the classical algorithm.

For the adiabatic quantum search, a Hamiltonian has to be  constructed which has the corresponding effect: 
$H(t)=f(t)H_0+g(t)H_1$  with the initial Hamiltonian
\begin{equation}
H_0=I-|\Psi_0\rangle\langle \Psi_0|,
\end{equation}
 which has the ''haystack'' as its ground state, and the final Hamiltonian
\begin{equation}
H_1=I-|m\rangle\langle m|,
\end{equation}
 which has the marked state $|m\rangle$ as its ground state. In its simplest version, with linear functions $f(t)=1-t/T$ and $g(t)= t/T$, the adiabatic search has a running time of $O(N)$, i.e. no speed-up over the classical search \cite{AQC}. This result can be improved by applying the adiabaticity condition locally, with respect to time, instead of globally ( to the entire time interval): The Hamiltonian can change faster when the gap between ground state and first excited state is large, which is the case at the beginning and the end of the evolution.  Only around $g_{min}$ it needs to change slowly. Adjusting the evolution rate with nonlinear functions $f(t)$ and $g(t)$  gives a running time of 
 $T_{min}= O(\sqrt{N})/\epsilon$ \cite{Roland0423}, corresponding to Grover's result. Recently it has been studied how  a temporary increase in energy for the oracle term, given by the function $g(t)$, can increase the speed-up further.  For a large number $n$ of qubits, the time $T$ for the computation is bounded below by \cite{Das2003} 
\begin{equation}
\frac{1}{\hbar}\int_0^T g(t)dt \geq\frac{k\sqrt{N}}{4} ,
\end{equation}
where $k=O(1)$. This result suggests that the action associated with the oracle term can be considered a resource required for adiabatic search. In the next section, I will study if entanglement can be considered a resource for adiabatic quantum search as well. 

\section{Entanglement in a two-qubit toy model}

\begin{figure}
\includegraphics[width=6cm]{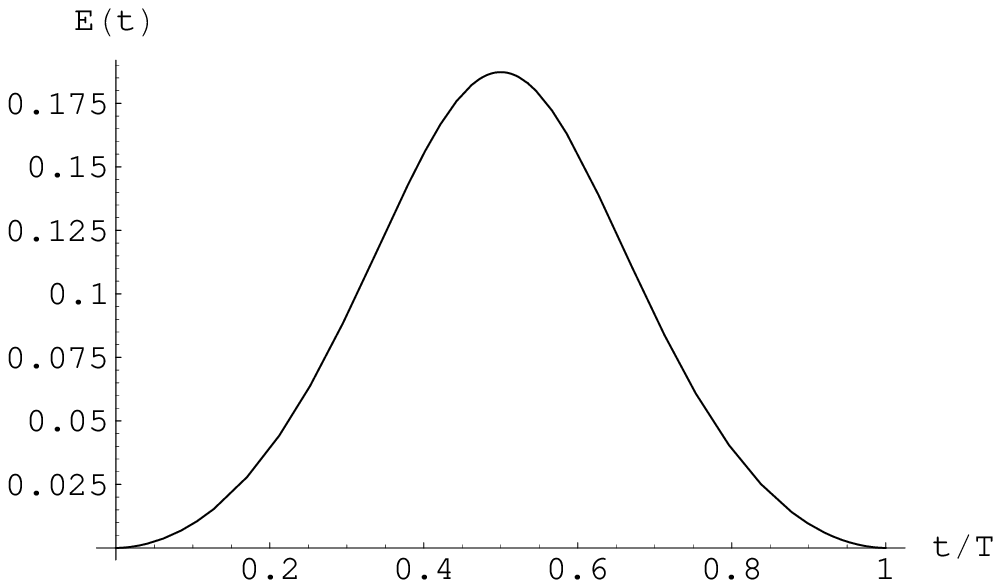}
\includegraphics[width=6cm]{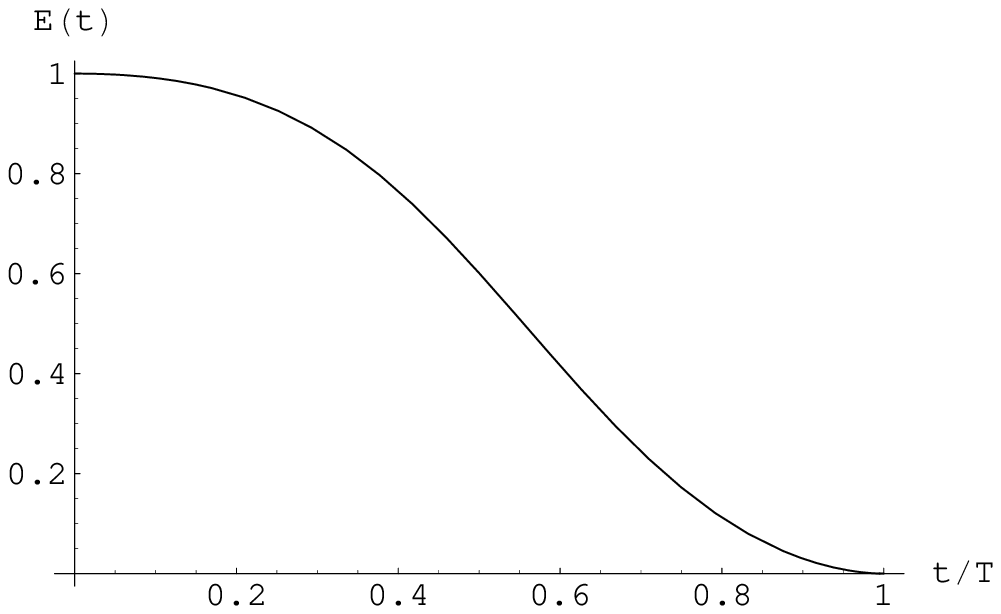}
\caption{Entropy of entanglement for the equally weighted superposition as initial state (left) and for a Bell state as initial state (right).}
\label{EqualBell}
\end{figure}

\begin{figure}
\includegraphics[width=6cm]{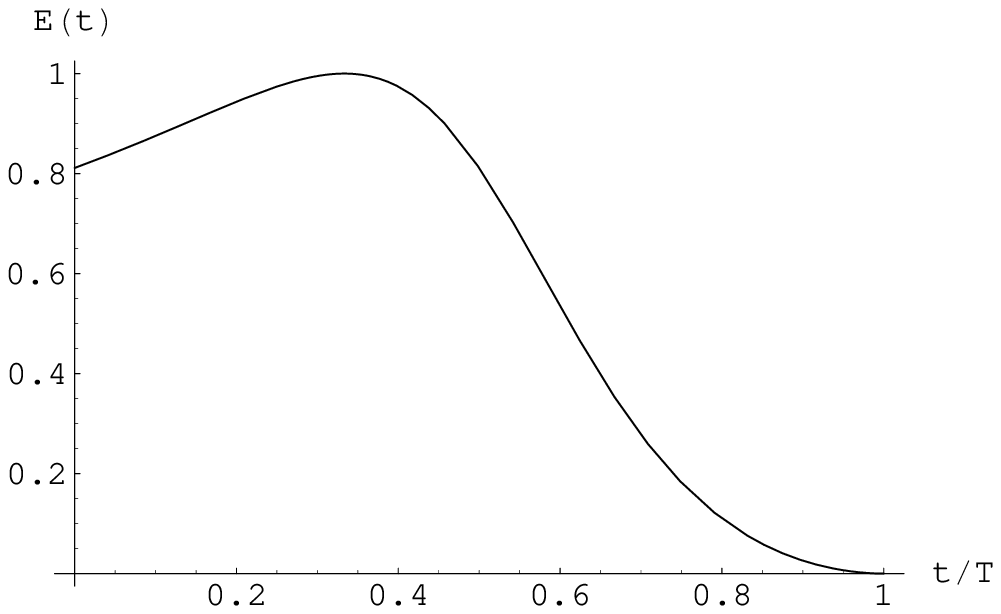}
\includegraphics[width=6cm]{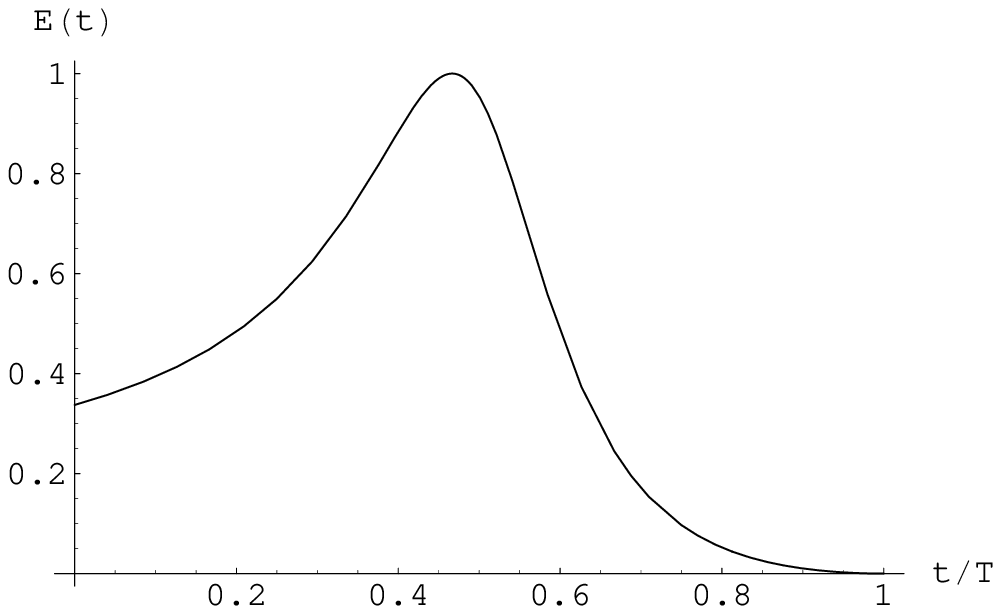}
\caption{Entropy of entanglement for two non-maximally entangled states as initial states.}
\label{nonmaxx}
\end{figure}

For the investigation of entanglement in adiabatic quantum search, I consider the simplest ''quantum computer'' possible, a system of two qubits. The computational basis states for this composite system are written as
 $|0\rangle\otimes |0\rangle =  |0\rangle |0\rangle =  |00\rangle , |01\rangle , |10\rangle , |11\rangle $. The initial state is 
\begin{equation}
|\Phi(0)\rangle =\frac{1}{\sqrt{4}}(|00\rangle+|01\rangle+|10\rangle+|11\rangle),
\end{equation}
and the marked state is taken to be $|00\rangle$. 
From these states, the time-dependent Hamiltonian is constructed as described in the previous section. Since for a quantum computer of this size there is no speed-up over classical computation, I can also use the simplest time dependence of the functions $f(t)$ and $g(t)$.  From the Hamiltonian, the lowest energy state is  calculated to be 
\begin{equation}
 |E_-(t)\rangle =\frac{1}{\sqrt{3(1+3(\frac{g(t)}{E_-(t)}-1)^2)}}
 \left(3(\frac{g(t)}{E_-(t)}-1),1,1,1\right),
\end{equation}
where $E_-=1/2(1-\sqrt{3(t/T)^2-3t/T+1})$ is the lowest energy eigenvalue.
Note that this corresponds to the general result (9) in \cite{Das2003} for the case of $n=2$ qubits and the functions $f(t)=1-t/T$ and $g(t)-t/T$.
 In the adiabatic approximation considered here, the systems stays in this instantaneous ground state throughout its evolution, which allows one to calculate the entropy of entanglement in a straightforward way: from the coordinates of the state in the computational basis, $c_0(t),c_1(t),c_2(t),c_3(t)$, I calculate the eigenvalues of its reduced density matrix as
\begin{equation}
  \mu_{\pm}(t)=\frac{1}{2}(1\pm\sqrt{1-4C^2(t)})
\end{equation}
with $C^2(t)=|c_0(t)c_3(t)-c_1(t)c_2(t)|^2$, the concurrence as a function of time. 
The entropy of entanglement for the ground state as a function of time is calculated from these eigenvalues as
$ \cal{E}  = -(\mu_+\log\mu_+ + \mu_-\log\mu_-)$.
The result is shown in Fig. \ref{EqualBell} (left).
Although both the initial and the final state are separable, the entanglement is non-zero  throughout the running time of the algorithm. Is this entanglement necessary for the algorithm to work, and to provide the speedup, or is it just a byproduct of the time evolution? For a more systematic approach, I vary the initial state, thereby modifying the search to one in which not all items are present with the same probability. (Note that the overlap of the initial and the marked state has to be nonzero for the search to succeed). A different initial state has, in general, a value for the entanglement different from the equally weighted superposition. It also leads to a different Hamiltonian, which induces different interactions in the system and therefore a different entanglement during the algorithm. Fig. \ref{nonmaxx} shows two typical results for  the entropy of entanglement for two initial states with non-maximal entanglement. 
In either case, the entanglement  reaches its maximum value of one during the evolution before going down to zero. In the second case, the initial state yields a Hamiltonian with  the same spectrum as the equally weighted superposition, and the same $T_{min}$; only the initial entanglement is different. This would support the assumption that the running time of the algorithm is correlated with the spectrum of the Hamiltonian, as the adiabatic theorem suggests, and not necessarily with the entanglement.  The entanglement for a search starting in a Bell state is shown in Fig. \ref{EqualBell} (right), starting at the maximum value one and going down to zero for the final state.

These four cases represent the results found when varying the initial state systematically. The entanglement either approaches the value zero monotonically, which seems to be the case for initial states with $c_0\geq c_3$, i.e. states ''close'' to the marked state, or it reaches a maximum during the running time. The value of this maximum seems to be independent of the initial entanglement, but the difference between maximum and initial entanglement seems to be larger for states with $c_3\gg c_0$. This would imply that for states ''further away'' from the marked state, a larger amount of entanglement is created during the search. 

In the cases studied, no correlation between the minimum energy gap $g_{min}$ and the initial entanglement could be found. Since the gap is correlated with the running time through the adiabatic theorem, this would also support the conclusion that entanglement is not the crucial resource for this algorithm.

\section{Discussion}

From the results presented in the previous section, no clear statement about the relevance of entanglement for the adiabatic quantum search algorithm can be made. This is not too surprising, since we know that the speed-up in Grover's search algorithm comes from the overlap of the initial and the marked state, $1/N=\langle\psi | m\rangle$. The relevance of this overlap, or the fidelity, for the algorithm should be investigated further \cite{prep}. Since $N$ is given by the dimension of the Hilbert space, it seems plausible that the Hilbert space structure is a crucial resource for the search algorithm, allowing a kind of parallel computation. But a caveat seems necessary here: The Hilbert space dimension is related to the dimension of the maximally entangled subspace \cite{Hayden}, and so, in a way, to the number of available entangled states. This could explain why there seems to be a contradiction between the results obtained here for a small system, and the results for large $N$ found in \cite{Osborne} for the standard Grover search: the more entangled the initial state is, the less well the algorithm performs. In that paper, the quality of performance is not the running time, but the probability for actually finding the marked state, which might not be comparable with the adiabatic running time. But the study of entanglement for systems with large $N$ in the adiabatic approach is certainly a future goal and will partially be addressed in \cite{prep}.
We plan to report on a larger, more systematic study of entanglement and fidelity during the search and other algorithms in \cite{prep}. Using numerical methods to obtain more extensive data, we will also attempt a more detailed interpretation of the resources used in adiabatic quantum computation.
From studying structured adiabatic quantum search \cite{structured} for large $N$, we know that a more structured, i.e. more local Hamiltonian has a shorter running time than a Hamiltonian including interactions between all subsystems. Since entanglement is created (and destroyed) by interactions, this would support the assumption  that for large systems, entanglement slows down the algorithm.  A better understanding of entanglement would generally be helpful for the understanding of the dynamical evolution of composite, interacting many-particle systems.

Acknowledgements:\\
It is a pleasure to thank Gabor Kunstatter, Randy Kobes, Meg Carrington and Todd Fugleberg for discussions at various stages of this work.

\end{document}